\documentclass[prl,twocolumn,amsmath,amssymb,superscriptaddress]{revtex4}

\usepackage{graphicx}
\usepackage{dcolumn}
\usepackage[ansinew]{inputenc}
\usepackage{bm}
\usepackage{amsfonts}
\usepackage{amssymb}

\newcommand{\ket}[1]{|#1\rangle}

\begin{document}

\title{Coherent manipulation of Bose-Einstein condensates with state-dependent microwave potentials on an atom chip}

\author{Pascal Böhi}
\author{Max F. Riedel}
\author{Johannes Hoffrogge}
\affiliation{Max-Planck-Institut für Quantenoptik und Sektion Physik der Ludwig-Maximilians-Universität, 80799 München, Germany}
\author{Jakob Reichel}
\affiliation{Laboratoire Kastler Brossel, ENS/UPMC-Paris 6/CNRS, 24 rue Lhomond, F-75005 Paris, France}
\author{Theodor W. Hänsch}
\affiliation{Max-Planck-Institut für Quantenoptik und Sektion Physik der Ludwig-Maximilians-Universität, 80799 München, Germany}
\author{Philipp Treutlein}\email[To whom correspondence should be addressed. E-mail: ]{treutlein@lmu.de}
\affiliation{Max-Planck-Institut für Quantenoptik und Sektion Physik der Ludwig-Maximilians-Universität, 80799 München, Germany}

\date{\today}

\begin{abstract}
Entanglement-based technologies, such as quantum information processing, quantum simulations, and quantum-enhanced metrology, have the potential to revolutionise our way of computing and measuring and help clarifying the puzzling concept of entanglement itself.
Ultracold atoms on atom chips are attractive for their implementation, as they provide control over quantum systems in 
compact, robust, and scalable setups. An important tool in this system is a potential depending on the internal atomic state.
Coherent dynamics in this potential combined with collisional interactions allows entanglement generation both for individual atoms and ensembles.
Here, we demonstrate coherent manipulation of Bose-condensed atoms in such a potential,
generated in a novel way with microwave near-fields on an atom chip. We reversibly entangle atomic internal and motional states, realizing a trapped-atom interferometer with internal-state labelling. Our system provides control over collisions in mesoscopic condensates, paving the way for on-chip generation of many-particle entanglement and quantum-enhanced metrology with spin-squeezed states.

\end{abstract}

\maketitle

Applications of ultracold neutral atoms in quantum information processing (QIP) \cite{diVincenzo00,Nielsen00}, quantum simulations \cite{Lloyd96,Jane03}, and quantum-enhanced metrology \cite{Dunningham06,Childs00} rely on the coherent control of internal states, motional states, and collisional interactions \cite{Chu02}.
Atom chips \cite{Fortagh07} are particularly attractive for the implementation of such entanglement-based technologies, because they provide versatile micropotentials for ultracold atoms in a compact, robust, and scalable setup \cite{Du04,Vogel06,Whitlock09}.
Coherent manipulation of internal \cite{Treutlein04} and motional \cite{Wang05,Schumm05,Guenther05,Guenther07,Jo07,Jo07b} states on atom chips has been demonstrated in separate experiments.
However, the combined coherent manipulation of internal and motional states with a state-dependent potential has not yet been achieved.
Such a manipulation is required for trapped-atom interferometry with internal-state labelling of the interferometer paths.
In combination with collisional interactions, it is a crucial ingredient for entanglement generation and at the heart of recently proposed schemes for atom chip quantum gates \cite{Calarco00,Treutlein06b} and spin-squeezing \cite{Li08b}.

\begin{figure}[tbh]
    \centering
        \includegraphics[width=0.4\textwidth]{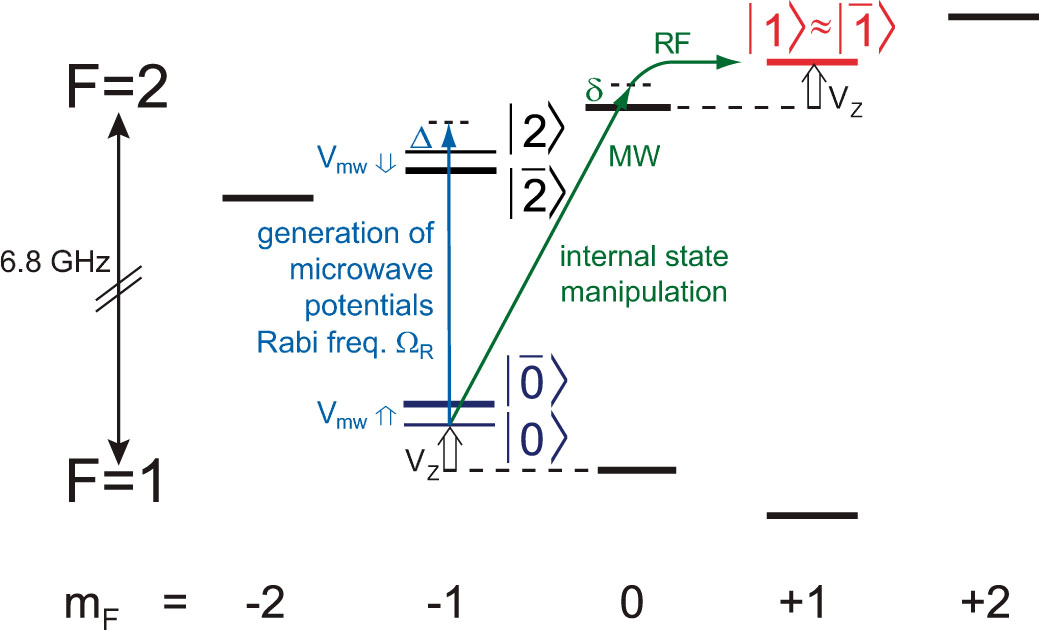}
    \caption{\textbf{Coherent internal-state manipulation and generation of state-dependent microwave potentials.}
    Hyperfine structure of the $^{87}$Rb ground state in the static magnetic field of the microtrap. The ``qubit'' or ``clock'' states $\ket{0}$ and $\ket{1}$ experience nearly identical Zeeman energy shifts $V_Z$. For internal-state manipulation of the atoms, the two-photon transition $\ket{0}\leftrightarrow \ket{1}$ is resonantly coupled, with detuning $\delta$ from the intermediate state $|F=2,m_F=0\rangle$. To generate state-dependent potentials, a microwave near-field is used, which couples $\ket{0}$ to the auxiliary state $\ket{2}$ with Rabi frequency $\Omega_R$ and detuning $\Delta$. The resulting dressed state $\ket{\bar 0}$ is shifted in energy by $V_\mathrm{mw}$ with respect to $|0\rangle$. State $\ket{1}$ is nearly unperturbed by the microwave near-field, because all transitions connecting to $\ket{1}$ are far off resonance.
     }
    \label{fig:RbTerms}
\end{figure}

For coherent internal-state manipulation on atom chips, the ground state hyperfine levels $|0\rangle\equiv |F=1,m_F=-1\rangle$ and $|1\rangle\equiv |F=2,m_F=1\rangle$ of $^{87}$Rb are an ideal choice (see Fig.~\ref{fig:RbTerms}). Both states are magnetically trappable, have nearly identical magnetic moments, and thus possess excellent coherence properties. This makes them ideal ``qubit'' or ``clock'' states compatible with magnetic traps. Internal-state manipulation of $\ket{0}$ and $\ket{1}$ with coherence lifetimes exceeding one second has been demonstrated at a few micrometers distance from the chip surface \cite{Treutlein04}.
Atoms in a coherent superposition of $\ket{0}$ and $\ket{1}$ can be entangled with each other via collisional interactions, provided that the interaction strength depends on the internal state.
For $\ket{0}$ and $\ket{1}$, the intra- and inter-state scattering lengths are nearly equal and no convenient Feshbach resonance exists which would allow tuning of the scattering lengths \cite{Marte02}.
Therefore, a state-dependent potential is required, which allows one to condition the interactions on the internal state by controlling the wave function overlap in a state-dependent way \cite{Jaksch99}.

Entanglement of atoms has been generated in this way in state-dependent optical lattice potentials \cite{Mandel03}.
The advantage of our system is that it is adaptable to a large range of atom numbers, ranging from large ensembles (as required for metrology with spin-squeezed states) over mesoscopic atom numbers (e.g.\ for experiments on entangled number states) in principle down to individual atoms \cite{Diener02,Mohring05,Colombe07} (as required for QIP). This derives from the fact that in chip-based magnetic traps, the atom number can be well adjusted through radio-frequency evaporative cooling \cite{Dudarev07} and trap frequencies and the trap aspect ratio can be easily tuned over a wide range (as opposed to the 1D situation of \cite{Widera08}). Furthermore, the atoms can be detected with high resolution both in space and in atom number for each internal state.

\section{Microwave near-field potentials}
In the experiments reported here, acting on our previous proposal \cite{Treutlein04,Treutlein06b}, we use microwave near-fields on an atom chip 
to generate state-dependent potentials for $\ket{0}$ and $\ket{1}$.
Microwave potentials derive from magnetic dipole transitions between the hyperfine states of the atomic ground state \cite{Agosta89}. The states are dressed by the microwave coupling, which gives rise to state-dependent energy shifts, because the Rabi frequency and detuning of the microwave are different for the different Zeeman sublevels. In a spatially varying microwave field, this results in a state-dependent potential landscape. Microwave potentials generated by far-field radiation were already studied in the 1990s \cite{Agosta89,Spreeuw94}. Hundreds of kilowatts of circulating microwave power inside a cavity were necessary, because the centimeter wavelength of the microwave prevents tight focussing and thus limits the attainable potential gradients.
Using microwave near-fields generated by micrometer-sized waveguides on atom chips, it is possible to realize much stronger gradients with only milliwatts of power.
This is because near-field gradients do not depend on the wavelength, but instead on the transverse waveguide dimensions and the distance from the waveguide. In addition, this allows tailoring of the potentials on the micrometre scale.
Recently, radio-frequency fields were used to generate potentials on atom chips \cite{Hofferberth06}. By comparison, microwave potentials have the important advantage that the different transitions are split by the Zeeman effect, which we make use of to adjust the state-dependence of the potentials simply via the microwave frequency.
Compared with optical potentials \cite{Bloch05}, microwave near-field potentials have the advantage of negligible spontaneous emission, tailorability, and a compact, chip-based setup.

Figure~\ref{fig:RbTerms} indicates the ground state hyperfine levels and transitions of $^{87}$Rb relevant to our experiments.
A static magnetic field $\mathbf{B}(\mathbf{r})$ is applied to generate a static magnetic trapping potential $V_Z(\mathbf{r})=\mu_B B(\mathbf{r})/2$, where $\mu_B$ is the Bohr magneton. The same potential $V_Z$ is experienced by both states $\ket{0}$ and $\ket{1}$.
The atoms trapped in $V_Z(\mathbf{r})$ can be placed in any desired superposition of $\ket{0}$ and $\ket{1}$ by resonantly coupling the two-photon transition $\ket{0}\leftrightarrow \ket{1}$ with microwave and radio-frequency far-field radiation generated off-chip 
\cite{Treutlein04}. This two-photon drive is turned on only for internal-state manipulation.
In order to create the state-dependent potential, a second microwave with magnetic field $\mathbf{B}_\mathrm{mw}(\mathbf{r},t)=\tfrac{1}{2}[\mathbf{B}_\mathrm{mw}(\mathbf{r})e^{-i\omega t} + \mathbf{B}_\mathrm{mw}^*(\mathbf{r})e^{i\omega t}]$ is applied.
This is a microwave near-field with strong gradients, generated by an on-chip waveguide.
The frequency $\omega$ is chosen such that it primarily couples $\ket{0}$ to the auxiliary state $\ket{2}\equiv\ket{F=2,m_F=-1}$, with position-dependent Rabi frequency $\Omega_R(\mathbf{r})=-\sqrt{3/4}(\mu_B/\hbar) B_\parallel(\mathbf{r})$ and detuning $\Delta(\mathbf{r})=\omega-\omega_\mathrm{hfs}+\mu_B B(\mathbf{r})/\hbar$. Here, $B_\parallel = \mathbf{B}\cdot\mathbf{B}_\mathrm{mw} / B$ is the microwave field component along the local direction of $\mathbf{B}$ and $\omega_\mathrm{hfs}$ is the hyperfine splitting for $B=0$. All transitions other than $\ket{0}\leftrightarrow \ket{2}$ are much further off resonance and therefore only have minor effects.
The coupling results in a dressed state $\ket{\bar 0}$ which is shifted in energy by $V_\mathrm{mw}(\mathbf{r})$ with respect to $\ket{0}$, the overall potential seen by $\ket{\bar 0}$ is thus $V_{\ket{\bar 0}} = V_Z + V_\mathrm{mw}$.
By contrast, state $\ket{1}$ and its potential remain essentially unchanged, $\ket{\bar 1}\approx \ket{1}$ and $V_{\ket{\bar 1}} \approx V_{\ket{1}} = V_Z$, because the microwave is very far off resonance from all transitions connecting to this state.
In this way, the microwave near-field adds internal-state dependence to the potential in a controlled way.
In our experiments, we focus on the regime $|\Omega_R|^2 \ll |\Delta|^2$, where $\ket{\bar 0}$ contains only a small admixture of state $\ket{2}$. This is important because $\ket{2}$ has opposite magnetic moment, and a large admixture would spoil the good coherence properties of our state pair.
In this limit, $V_\mathrm{mw}(\mathbf{r}) \approx \hbar |\Omega_R(\mathbf{r})|^2/4\Delta(\mathbf{r})$ and $\ket{\bar 0} \approx \ket{0} + \left(\Omega_R(\mathbf{r})/2\Delta(\mathbf{r})\right) \ket{2}$.
In the general case, we determine $\ket{\bar 0}$, $\ket{\bar 1}$, $V_{\ket{\bar 0}}$, and $V_{\ket{\bar 1}}$ using a numerical simulation that takes all hyperfine transitions into account, is valid for arbitrary detunings, and includes several other contributions to the potential (see Supplementary Information).

\section{Microwave Atom Chip}
\begin{figure}[tbhp]
    \centering
        \includegraphics[width=0.50\textwidth]{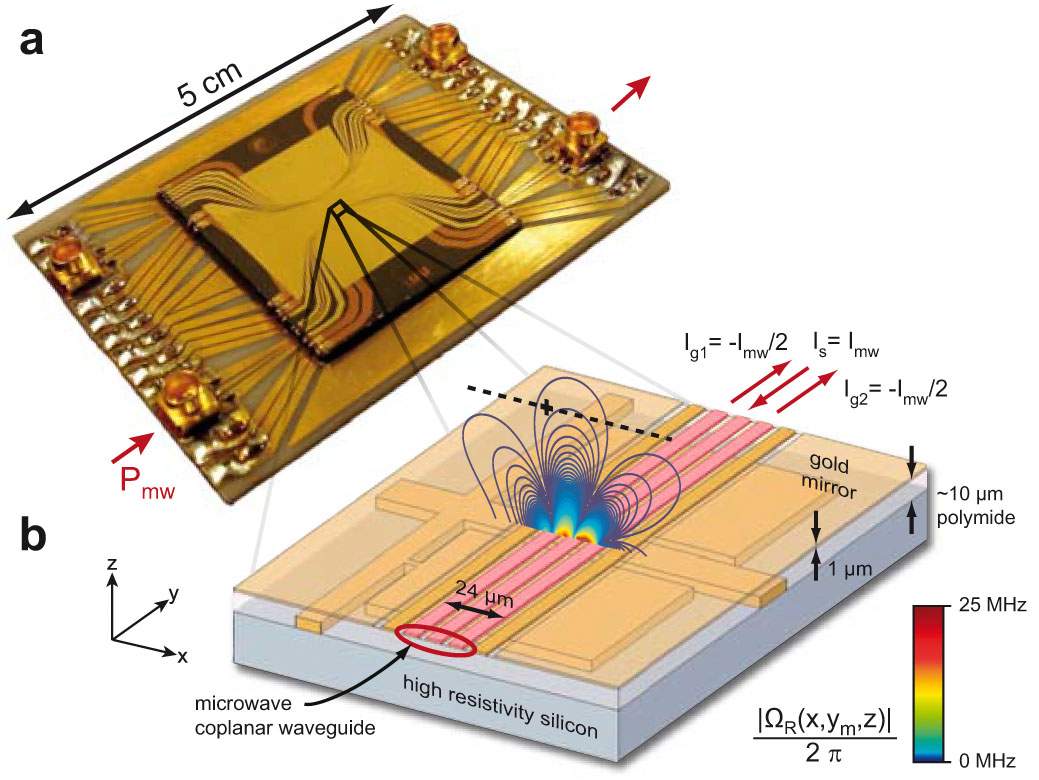}
    \caption{\textbf{Atom chip with microwave coplanar waveguides (CPWs).}
    {\bf a}, Photograph of the chip assembly. The Si experiment chip has two layers of gold wires separated by a thin polyimide layer, with CPWs on the upper layer. It is glued and wire bonded to an AlN carrier chip with a single gold layer, providing four microwave and $44$ DC ports.
    {\bf b}, Schematic close-up of the experiment region. The three central wires (red, cross section $6~\mu\mathrm{m}\times 1~\mu\mathrm{m}$ each) form a CPW. All wires (including the CPW) can carry stationary currents for the generation of static magnetic traps. The position $\mathbf{r}_m=(x_m,y_m,z_m)=(-12,0,44)~\mu$m of the minimum of the static trap $V_Z(\mathbf{r})$ is indicated by the black cross ($\mathbf{r}=0$ corresponds to the top surface of the wire in the center of the CPW). An ideal CPW mode has microwave current amplitudes $I_s=I_\mathrm{mw}$ and $I_{g1}=I_{g2}=-I_\mathrm{mw}/2$ as indicated. The field of these currents is used to calculate $\Omega_R(\mathbf{r})$ (see Methods). Equipotential lines of $|\Omega_R(x,y_m,z)|/2\pi$ are shown for $I_\mathrm{mw}=76$~mA (line spacing 70~kHz, for an enlarged version of the plot, see Supplementary Information). The asymmetry in $|\Omega_R(\mathbf{r})|$ with respect to $x=0$ is due to the spatial dependence of $\mathbf{B}(\mathbf{r})$ which gives rise to $V_Z(\mathbf{r})$.
\label{fig:chiplayoutschem}}
\end{figure}

We have built a novel type of multi-layer atom chip, see Fig.~\ref{fig:chiplayoutschem}a, with integrated coplanar waveguides (CPWs) for the generation of microwave near-fields in addition to wires for static magnetic fields \cite{TreutleinThesis08}.
The CPWs are tapered from millimetre size at the connectors to a few micrometres in the chip centre, with only small changes in impedance. This ensures high microwave transmission to the experiment region.
In the present work, we use the waveguide which is shown schematically in the close-up of Fig.~\ref{fig:chiplayoutschem}b.
We have simulated microwave propagation on this structure, including both layers of metallization and dielectrics (see Methods).
We find that at the position of the atoms, $\mathbf{B}_\mathrm{mw}(\mathbf{r})$ can be well approximated by the static field of the currents indicated in Fig.~\ref{fig:chiplayoutschem}b. This corresponds to assuming an idealised CPW mode with transverse electromagnetic fields \cite{Collin01} and neglecting effects of the other wires. The relative error in the spatial dependence of $\mathbf{B}_\mathrm{mw}(\mathbf{r})$ introduced by this approximation is about 10\%, comparable to other uncertainties in the microwave current distribution due to the wire bonds, CPW tapers, and chip connectors.

The experiments are performed in the following way.
Using a combination of stationary currents and external fields, we prepare a Bose-Einstein condensate (BEC) \cite{Ketterle99} in state $\ket{0}$ in a static magnetic microtrap.
Very stable current sources and magnetic shielding allow us to reproducibly prepare BECs with small atom number, typically $N=400\pm 21$ (see Methods).
The BEC can be precisely positioned in the near-field of the CPW by ramping appropriate wire currents and fields. For the experiments presented here, the BEC is transferred into a cigar-shaped harmonic trap $V_Z(\mathbf{r})$, with long axis along $x$, measured trap frequency $f_x=109$~Hz ($f_\perp=500$~Hz) in the axial (radial) direction, and measured static field in the trap center $B(\mathbf{r}_m)=3.23$~G, pointing along $x$. The position $\mathbf{r}_m$ of the minimum of $V_Z(\mathbf{r})$ is indicated in Fig.~\ref{fig:chiplayoutschem}b.
If the microwave on the CPW is turned on while the atoms are trapped in $V_Z(\mathbf{r})$, they see a strongly position-dependent Rabi frequency $\Omega_R(\mathbf{r})$, as shown in Fig.~\ref{fig:chiplayoutschem}b.
We now describe two experiments in which we use the resulting microwave potential $V_\mathrm{mw}(\mathbf{r})$ for state-dependent manipulation. 

\section{State-selective Splitting of a BEC}
\begin{figure}[tbh]
    \centering
        \includegraphics[width=0.5\textwidth]{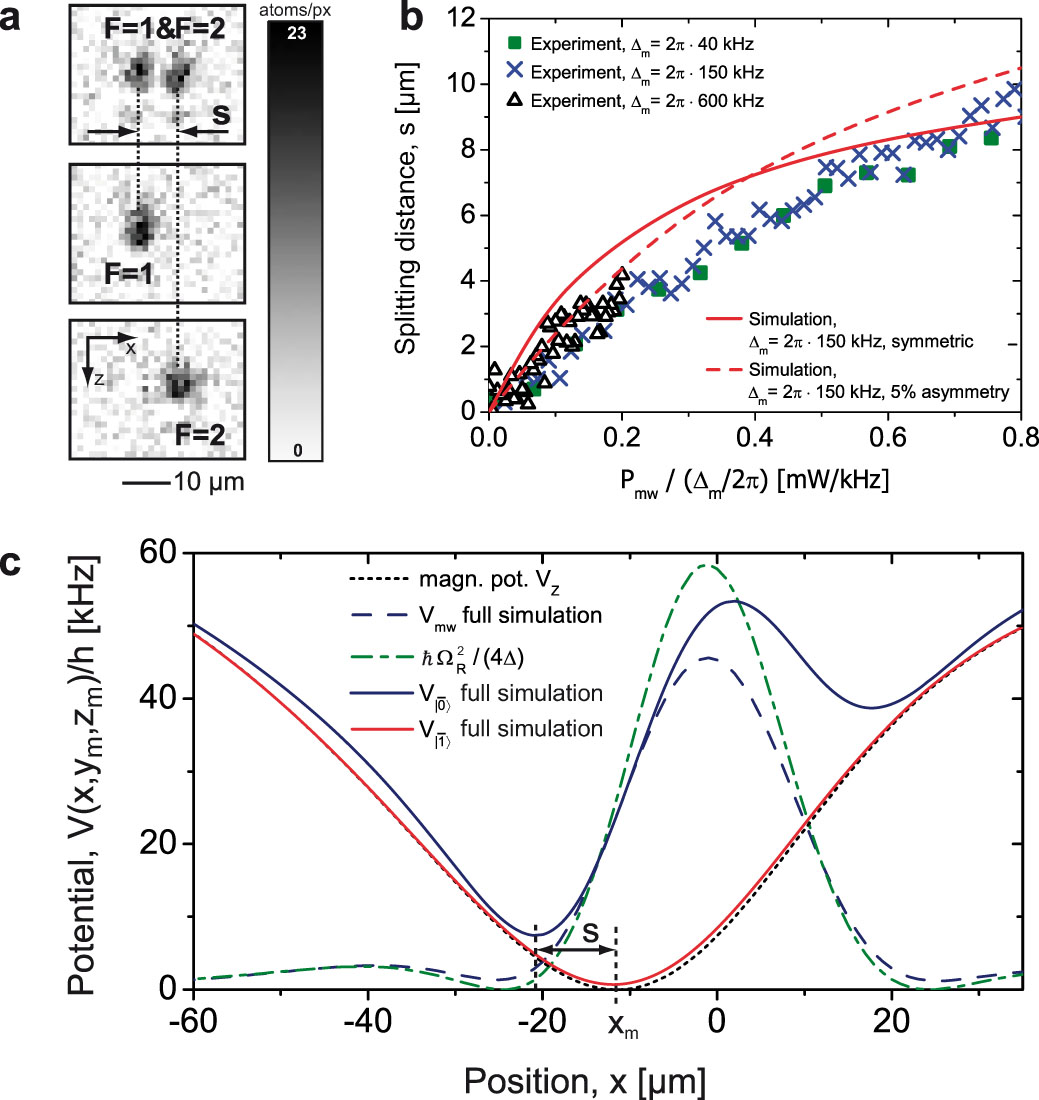}
    \caption{\textbf{State-selective splitting of the BEC.} \textbf{a}, Absorption images of the adiabatically split BEC ($P_\mathrm{mw} = 120$~mW, $\Delta_m\equiv \Delta(\mathbf{r}_m)=2\pi \times 150$~kHz). By imaging both hyperfine states simultaneously (top), only $F=1$ (middle), or only $F=2$ (bottom), the state-selectivity of the splitting is established.
    \textbf{b}, Measured splitting distance $s$ as a function of $P_\mathrm{mw}/\Delta_m$ for different values of $\Delta_m$ as indicated. The solid line is the result of a simulation assuming an ideal CPW mode ($I_{\mathrm{g1}}=I_{\mathrm{g2}}= -0.5\,I_s$), the dashed line assumes a slightly asymmetric mode ($I_{\mathrm{g1}}= -0.45\,I_s$, $I_{\mathrm{g2}}=-0.55\,I_s$). The remaining discrepancy can be attributed to our simple simulation not reproducing the actual microwave near-field distribution perfectly.
    \textbf{c}, Simulated potentials along the splitting direction (see dotted line in Fig.~\ref{fig:chiplayoutschem}b), for an ideal CPW mode and $I_\mathrm{mw}=76$~mA, corresponding to the parameters of (\textbf{a}). The potential minimum of $V_{\ket{\bar 0}}$ is shifted by the microwave, while $V_{\ket{\bar 1}} \approx V_Z$. The full microwave potential $V_\mathrm{mw}$ (dashed blue) and the approximation $V_\mathrm{mw} \approx \hbar |\Omega_R|^2/4\Delta$ for $|\Omega_R|^2\ll |\Delta|^2$ (dash dotted green) are shown in comparison.
    }
    \label{fig:VerschiebungNeueFalleComb}
\end{figure}

In the first experiment, we show state-selective adiabatic splitting of the BEC (Fig.~\ref{fig:VerschiebungNeueFalleComb}).
Before we turn on the microwave potential, we prepare the atoms in $V_Z$ in the coherent superposition of internal states $\tfrac{1}{\sqrt{2}}(\ket{0}+\ket{1})$ by applying a $\tfrac{\pi}{2}$-pulse of $170~\mu$s duration on the two-photon transition. Right after this pulse, which is fast compared with the trap oscillation periods, the motional wave functions of $\ket{0}$ and $\ket{1}$ overlap completely. Then, within 150~ms, we smoothly ramp up the microwave power launched into the CPW to a final value of $P_\mathrm{mw}= 120$~mW, at fixed detuning $\Delta_m \equiv \Delta(\mathbf{r}_m) = 2\pi \times 150$~kHz. This corresponds to a ramp of $V_\mathrm{mw}$ which is adiabatic with respect to the dynamics of the internal state, ensuring population of only state $\ket{\overline{0}}$ but not $\ket{\overline{2}}$, as well as to the motion, allowing the BEC wave function to follow the potential.
At the end of the ramp, we switch off the combined static and microwave potentials within $0.3$~ms and image the atomic density distributions quasi in-situ, using state-selective absorption imaging \cite{Matthews98} to discriminate between $\ket{0}$ and $\ket{1}$. Figure~\ref{fig:VerschiebungNeueFalleComb}a shows images taken in this way.
We observe that the BEC is state-selectively split along $x$ by a distance $s=9.4~\mu$m, which is $3.9$ times the radius of each of the two trapped clouds \cite{MunozMateo07}.

The  splitting is due to the strong near-field gradient in $|\Omega_R(\mathbf{r})|$ around $\mathbf{r}=\mathbf{r}_m$. By comparison, the spatial dependence of $\Delta(\mathbf{r})$ is weak. Although $|\Omega_R(\mathbf{r})|$ has gradients of similar magnitude along the $x$ and $z$ axes (cf.\ Fig.~\ref{fig:chiplayoutschem}b), the spatial splitting is nearly one-dimensional because $f_{\perp}^{2} \gg f_{x}^{2}$.
Figure~\ref{fig:VerschiebungNeueFalleComb}b shows the measured $s$ as a function of $P_\mathrm{mw}/\Delta_m$ for different values of $\Delta_m$. The data points lie on top of each other as expected from the scaling $V_\mathrm{mw} \sim |\Omega_R|^2/\Delta \sim P_\mathrm{mw}/\Delta_m$ in the regime $|\Omega_R|^2 \ll |\Delta|^2$.
The maximally applied $P_\mathrm{mw}= 120$~mW corresponds to $\Omega_R(\mathbf{r}_m) = 2\pi\times 122$~kHz, which we measure independently by driving resonant Rabi oscillations with the microwave near-field. Note that for $\Delta >0$, the repulsive microwave potential pushes state $\ket{\bar 0}$ into regions where $\Omega_R(\mathbf{r}) \ll \Omega_R(\mathbf{r}_m)$ so that $|\Omega_R|^2 \ll |\Delta|^2$ is satisfied.

The observed splitting is reproduced by our simulation of $V_{\ket{\bar 0}}$ and $V_{\ket{\bar 1}}$, where the microwave current amplitude $I_\mathrm{mw}$ is calibrated using the measured $\Omega_R(\mathbf{r}_m)$ (see Methods).
Figure~\ref{fig:VerschiebungNeueFalleComb}c shows a slice through the simulated potentials along the splitting direction, assuming an ideal CPW mode.
In agreement with the experiment, the simulation shows that we can selectively displace the wave function of state $\ket{\bar 0}$ with the microwave potential gradient, while state $\ket{ \bar 1}$ is nearly unaffected.
Figure~\ref{fig:VerschiebungNeueFalleComb}b shows that the observed dependence of $s$ on $P_\mathrm{mw}/\Delta_m$ can be reproduced even better by assuming a slightly asymmetric CPW mode, which can arise due to asymmetries of the tapers and wire bonds.

\section{Demonstration of fully coherent operation}
\begin{figure}[tbh]
   \centering
   \includegraphics[width=0.47\textwidth]{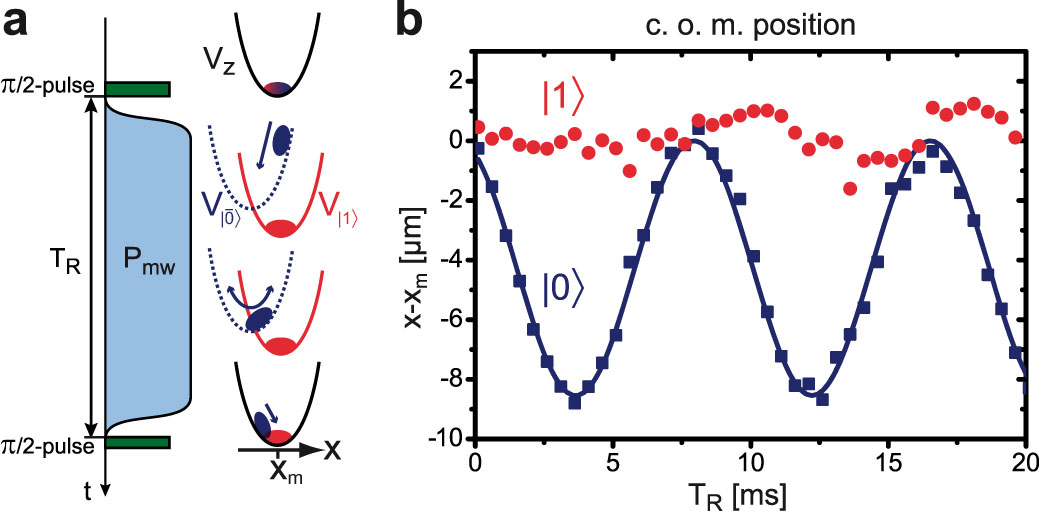}
   \caption{\textbf{Dynamical splitting and recombination scheme used for BEC interferometry.} \textbf{a}, Timing sequence of the interferometer. In between the two $\tfrac{\pi}{2}$-pulses of a Ramsey sequence on the $\ket{0} \leftrightarrow \ket{1}$ transition, the microwave on the CPW is pulsed on for a duration $T_R$, resulting in a sudden displacement of the potential minimum of $V_{\ket{\bar 0}}$. This sets the wave function of state $\ket{\bar 0}$ into oscillation.
   \textbf{b}, Oscillation of the atoms, recorded with the sequence of (\textbf{a}), but with the second $\tfrac{\pi}{2}$-pulse omitted. The centre-of-mass position of the atoms at the end of the sequence is shown as a function of $T_R$. State $\ket{0}$ oscillates while state $\ket{1}$ remains initially at rest. Each time the wave functions overlap in the trap, energy is transferred between the states.
   }
   \label{fig:insitu_position}
\end{figure}

In the second experiment, we demonstrate fully coherent control of the atoms by performing trapped-BEC interferometry with internal-state labelling of the interferometer arms.
Our interferometer consists of a Ramsey $\tfrac{\pi}{2}$-$\tfrac{\pi}{2}$ sequence on the $\ket{0} \leftrightarrow \ket{1}$ transition in combination with state-dependent splitting and recombination of the motional wave functions. We use a non-adiabatic splitting and recombination scheme, see Fig.~\ref{fig:insitu_position}a, which is motivated by the sequence required for the atom chip quantum gate proposed in \cite{Calarco00}. By choosing $\Delta_m = 2\pi\times 600$~kHz we ensure that the admixture of state $\ket{2}$ is small enough so that decoherence due to magnetic field noise is not a problem on the time scale of our experiment.
After the first $\tfrac{\pi}{2}$-pulse, the microwave on the CPW is switched on within $50~\mu$s to $P_\mathrm{mw}=120$~mW, which corresponds to a sudden displacement of the potential minimum for state $\ket{\bar 0}$ by $4.3~\mu$m.
After a variable delay, we switch off the microwave within $50~\mu$s, followed by the second $\tfrac{\pi}{2}$-pulse and state-selective detection to determine the number of atoms $N_0$ ($N_1$) in state $\ket{0}$ ($\ket{1}$). The time between the $\tfrac{\pi}{2}$-pulses, $T_R$, corresponds to the overall time the microwave was turned on.
In this scheme, the switching of $V_\mathrm{mw}$ is adiabatic with respect to the internal-state dynamics, but fast compared to the trap oscillation periods.
The wave function of state $\ket{\bar 0}$ is thus set into oscillation in the shifted potential $V_{\ket{\bar 0}}$. We can record these oscillations by varying $T_R$ and imaging the atoms without applying the second $\tfrac{\pi}{2}$-pulse, see Fig.~\ref{fig:insitu_position}b. The wave function of $\ket{\bar 0}$ oscillates with a peak-to-peak amplitude of $8.5~\mu$m and a frequency of $\bar f_x = 116$~Hz, which is the trap frequency of $V_{\ket{\bar 0}}$ along $x$. Periodically, it comes back to its initial position, approximately when $T_R$ is an integer multiple of $\bar f_x^{-1} = 8.6$~ms. At these times, it overlaps with the wave function of state $\ket{1}$.
Note that due to collisions, state $\ket{1}$ starts to oscillate as well.

\begin{figure*}[tbh]
    \centering
    \includegraphics[width=0.9\textwidth]{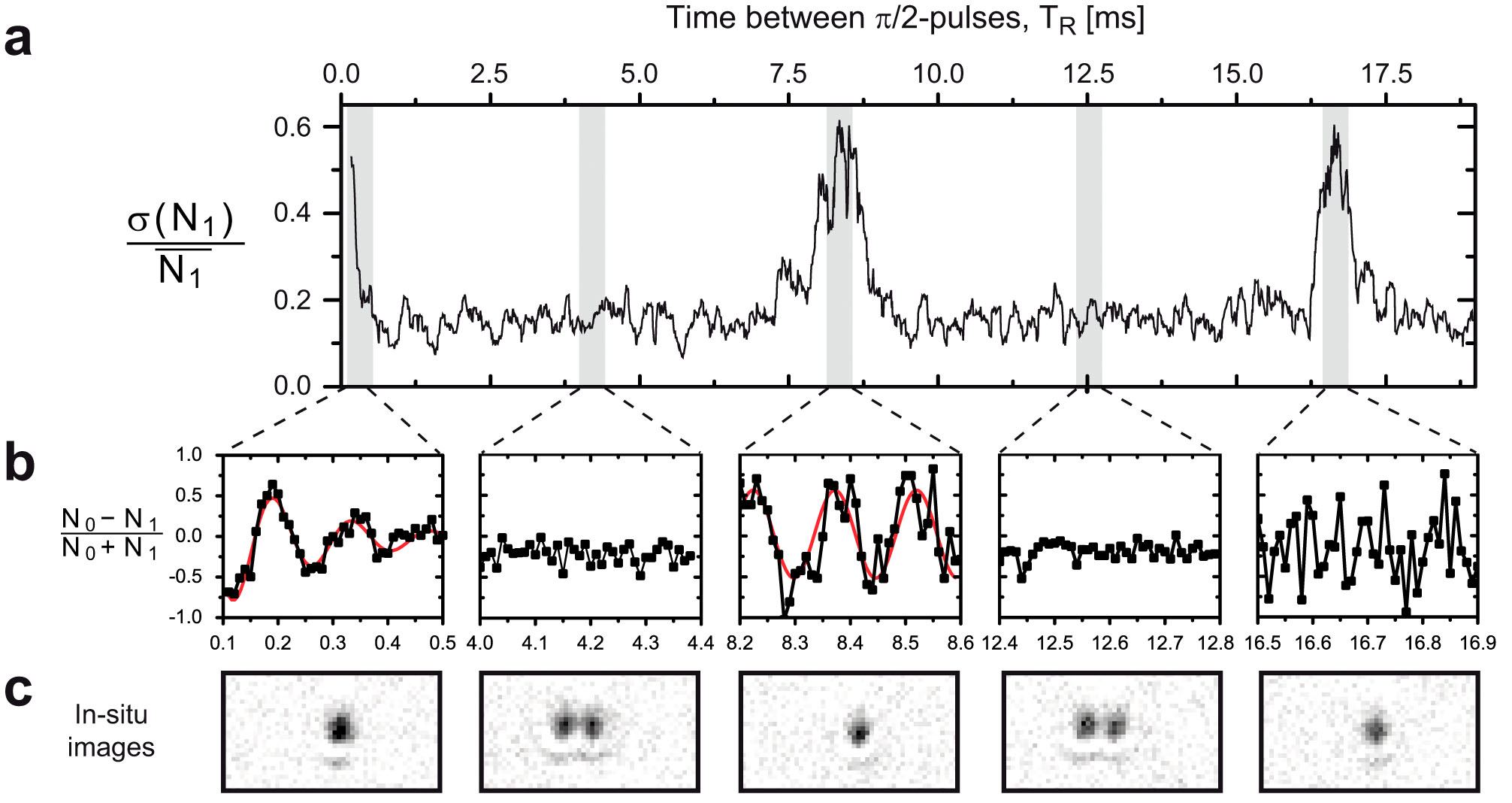}
    \caption{\textbf{Periodic recurrences of Ramsey interference contrast in the BEC interferometer.} The contrast of the Ramsey fringes on the $\ket{0}\leftrightarrow \ket{1}$ transition is modulated due to the periodic splitting and recombination of the motional wave functions.  \textbf{a}, As a measure of the wave function overlap, we show $\sigma(N_1)/\bar N_1$ as a function of $T_R$, where $\sigma(N_1)$ is the standard deviation and $\bar N_1$ the mean of $N_1$ obtained from a running average over a time interval $\left[T_R-75~\mu\mathrm{s}, T_R+75~\mu\mathrm{s}\right]$, corresponding to one period of the Ramsey fringes. The width of the recurrence is influenced by nonlinear wavefunction dynamics due to mean-field interactions.
    \textbf{b}, Corresponding Ramsey fringe data for selected values of $T_R$. Each data point is determined from two consecutive runs of the experiment, in which either $N_1$ or $N_0$ is detected. The surplus of atoms in state $\ket{1}$ at times when the contrast has vanished (second and fourth graph) is probably due to small intensity gradients of the microwave used to drive the two-photon transition, caused by near-field effects due to the microstructured surface.
    \textbf{c}, In-situ images of the atomic density distribution of $\ket{0}$ and $\ket{1}$, for $T_R$ corresponding to the centre of the windows in (\textbf{b}).
    }
    \label{fig:SSEbyMEANCombined}
\end{figure*}

If we apply both $\tfrac{\pi}{2}$-pulses and vary $T_R$, we observe Ramsey interference fringes, see Fig.~\ref{fig:SSEbyMEANCombined}. The interference contrast is modulated by the wave function overlap of the two states and thus periodically vanishes and reappears again due to the oscillation of state $\ket{\bar 0}$. As a measure of the wave function overlap we plot $\sigma(N_1)/\bar N_1$ as a function of $T_R$, where $\sigma(N_1)$ is the standard deviation and $\bar N_1$ the mean of $N_1$ obtained from a running average over one period of the Ramsey fringes, see Fig.~\ref{fig:SSEbyMEANCombined}a. This measure of the overlap has the advantage that it is largely insensitive to noise on the Ramsey fringes. Corresponding fringe data and in-situ images of the atoms at specific times $T_R$ are shown in Figure~\ref{fig:SSEbyMEANCombined}b+c. Precisely at the time when state $\ket{\bar 0}$ has performed a full oscillation in $V_{\ket{\bar 0}}$ a sharp recurrence of the contrast is observed. The recurrence of the interference proves that the combined evolution of internal and motional state is coherent.
The high contrast of the first recurrence shows that the collisional interactions between the atoms observable in Fig.~\ref{fig:insitu_position}b lead only to a small distortion of the wave functions. Wave function distortion can be reduced to negligible levels by optimal control of the splitting process as discussed in \cite{Treutlein06b}.

For the second (and subsequent) recurrences we observe substantial noise on the Ramsey fringe data. By contrast, when we take Ramsey fringes without splitting the BEC, comparable noise is visible only for $T_R$ beyond several hundred milliseconds. We have analysed technical fluctuations of the potentials, however, we find that they can account only for about 30\% of the observed noise level (see Methods). The additional noise could be due to spin squeezing in the BEC, as suggested in~\cite{Li08b}, which in the present $\tfrac{\pi}{2}$-$\tfrac{\pi}{2}$ sequence would show up as increased phase noise.
This effect will be studied in future experiments.
It suggests that our system can be used to tune interactions in a state-dependent way for atoms such as $^{87}$Rb that do not have convenient Feshbach resonances. This could lead to the realisation of many-particle entangled states such as spin-squeezed states \cite{Li08b} or Schrödinger cat states \cite{Sorensen01,Micheli03}.

\section{Applications of microwave near-fields}
Internal-state labelling of the interferometer paths \cite{Borde89,Berman97}, as demonstrated here, offers several advantages for trapped-atom interferometry.
Compared with state-insensitive beam splitters which operate by ramping up a barrier in the potential, the splitting and recombination can be controlled much more accurately by driving the internal-state transition. The interferometer paths can be closed inside the trap without the excitation of solitons which subsequently decay into vortices \cite{Jo07b,Scott08}. Furthermore, readout is greatly simplified, because it does not require spatially resolving interference patterns. Instead, only $N_0$ and $N_1$ have to be determined, which can be done with high accuracy. Many-body effects in the interferometer could either be suppressed by adjusting the trap frequencies for operation at lower density, or used beneficially to increase measurement precision with spin-squeezed states.

The oscillation of state $\ket{\bar 0}$ results in periodic entanglement and disentanglement of internal and motional states of the atoms. This mechanism is at the heart of the quantum phase gate proposed in \cite{Calarco00,Treutlein06b}. The gate can be viewed as two state-selective interferometers next to each other, each containing a single atom, where the oscillating states collide and pick up a phase shift, resulting in entanglement between the two atoms.
We envisage that the gate can be realised by combining the microwave potentials demonstrated here with techniques for single atom control \cite{Diener02,Mohring05} based on the recently demonstrated optical fibre cavities on atom chips \cite{Colombe07}.
As an alternative to single-atom operation, it is interesting to study whether two BEC interferometers next to each other could be used in a similar way to generate many-particle entanglement, and whether this entanglement could be used for QIP.

Beyond QIP, many-particle entanglement, and atom interferometry, microwave near-fields are a useful new tool for a variety of experiments. Chip-based atomic clocks that benefit from the large Rabi frequencies, tailored field configurations, and temporal stability of microwave near-fields are currently being set up \cite{Rosenbusch09}.
Microwave near-field potentials could be utilised to trap neutral atoms in internal states which cannot be trapped with static fields \cite{Agosta89,Spreeuw94}, or to realise electrodynamic traps for ultracold molecules or electrons.
An extension to multiple microwave frequencies seems promising and is technically straightforward.

{\small
\section{Methods}
\subsection*{Experimental Setup}
Our compact atom chip setup is surrounded by a $\mathrm{\mu}$-metal shield which reduces ambient magnetic field fluctuations to $0.2$~mG r.m.s. This, together with the use of very stable current sources, 
results in a high stability of the potentials allowing the preparation of BECs with mesoscopic atom number.

We start our experiments with a mirror-magneto-optical trap \cite{Haensel01a} followed by optical molasses to precool about $10^7$ $\mathrm{^{87}Rb}$ atoms to a temperature of $10\,\mathrm{\mu K}$. After optical pumping to the $|0\rangle\equiv |F=1,m_F=-1\rangle$ state the atoms are magnetically trapped, gradually moved closer to the chip surface and evaporatively cooled by applying three radio-frequency (RF) ramps which leads to BECs containing about $N=3000$ atoms.
We then cut further into the condensate with the RF knife in order to prepare small, pure BECs with well defined atom number, and transfer the BEC into the experiment trap (see main text).

For internal-state manipulation we use microwave radiation from an off-chip horn and RF radiation from an off-chip antenna. The microwave is blue detuned by $\delta/2\pi = 280$~kHz from the transition $\ket{1,-1} \leftrightarrow \ket{2,0}$. The duration of a $\frac{\pi}{2}$-pulse between $\ket{0}$ and $\ket{1}$ is $170~\mu$s. In the static magnetic trap $V_Z$, we observe Ramsey fringes with a contrast $>90\%$ and a coherence lifetime of $\approx 1$~s.

We detect the atoms state-selectively by absorption imaging \cite{Matthews98}, after a time-of-flight (TOF) of either $0.1$~ms or $4.0$~ms. The short TOF allows for a quasi-in-situ determination of the BEC position but diffraction of the imaging beam on the chip wires prevents an accurate determination of $N$.
After the longer TOF, on the other hand, the atoms are sufficiently far from the chip surface so that they can be imaged with the imaging beam aligned parallel to the chip. This results in a more accurate determination of $N$. We calibrate the detected atom number following \cite{Reinaudi07}.

\subsection*{Microwave field simulation}

We simulate the microwave current distribution in the center region of the chip using the software package \textit{Sonnet}.
For the simulation, we assume that an ideal coplanar waveguide mode is injected into the CPW, with currents $I_s(t) = I_\mathrm{mw}\cos(\omega t)$ on the center wire and $I_\mathrm{g1}(t)=I_\mathrm{g2}(t)=-(I_\mathrm{mw}/2)\cos(\omega t)$ on the ground wires (Fig.~\ref{fig:chiplayoutschem}b). The simulation result shows induced currents in the additional wires next to the CPW grounds with an amplitude of 2.3\% of $I_{\mathrm{mw}}$, in-phase with $I_s(t)$, and induced currents in the lower wire layer which essentially mirror the currents in the CPW, with $\approx 20$\% amplitude.
For distances from the chip much larger than the CPW wire dimensions, such as our typical trap positions ($z_m \approx 40\,\mathrm{\mu m}$), we find that neglecting all induced currents and assuming a homogeneous current distribution in the cross section of the three CPW wires, the splitting distance $s$ as a function of $P_\mathrm{mw}/\Delta_m$ changes by less than 10\% compared with the full simulation. This error is of the same order of magnitude as uncertainties in the CPW excitation. For simplicity, we therefore neglect induced currents and approximate $\mathbf{B}_\mathrm{mw}(t)$ at any given time $t$ with the static field of currents $I_s(t)$, $I_\mathrm{g1}(t)$, and $I_\mathrm{g2}(t)$ with homogeneous distribution in the wires. This approximation amounts to neglecting retardation and propagation effects of the microwave field.

Comparing simulations and measurements of the splitting distance $s$ as a function of $P_\mathrm{mw}$, we find that the observed characteristics can be best understood by assuming that the CPW is excited with a non-symmetric microwave current distribution on the two ground wires, with $I_{\mathrm{g1}}= -0.45\,I_{\mathrm{mw}}$ and $I_{\mathrm{g2}}=-0.55\,I_{\mathrm{mw}}$. The asymmetry can be understood as a partial coupling from the even coplanar microwave mode to the odd one \cite{Ghione87}. It could arise due to asymmetries in the bond wires and the curves of the CPW on the chip.

We calibrate the microwave current amplitude in the chip centre used in the simulation, $I_\mathrm{mw}$, to the microwave power launched into the chip in the experiment, $P_\mathrm{mw}$, by comparing the simulated and measured Rabi frequencies $\Omega_R(\mathbf{r}_m)$ at the position of the static trap minimum.

\subsection*{Noise on the Ramsey fringes}
While states $\ket{0}$ and $\ket{1}$ have identical magnetic moments at $B=3.23$~G, the magnetic moment of state $\ket{\bar 0}$ is different from that of $\ket{1}$ because of the admixture of state $\ket{2}$. This makes the dressed qubit state pair sensitive to magnetic field fluctuations. We directly measure the magnetic field sensitivity of the Ramsey fringes in Fig.~\ref{fig:SSEbyMEANCombined} by scanning $B(\mathbf{r}_m)$ in the experiment, and find that a static magnetic field change of $16$~mG leads to a phase shift of $2\pi$ at the time of the first recurrence ($T_R = 8.4$~ms). Inside the magnetic shielding, we measure r.m.s.\ fluctuations of $\delta B = 0.2$~mG, corresponding to r.m.s.\ phase fluctuations of $\delta\phi_B = 0.03\,\pi$ at $T_R = 8.4$~ms. Surface effects, such as loss and decoherence due to thermal magnetic near-field noise \cite{Fortagh07}, are negligible in the present experiments because of the thin metallic chip layers and relatively large atom-surface distances. We also measure the sensitivity of the phase of the Ramsey signal on changes of $P_\mathrm{mw}$, and use it to estimate that the measured r.m.s.\ fluctuations of $\delta P_\mathrm{mw} = 20~\mu$W result in phase fluctuations of $\delta\phi_P = 0.01\,\pi$ at $T_R = 8.4$~ms. We furthermore include the quantum projection noise, $\delta\phi_{S}=1/\sqrt{N}=0.02\,\pi$, and the measured r.m.s.\ fluctuations in total atom number, $\delta N = 21$. Taken together, these effects can explain only about $30$\% of the noise level on the Ramsey fringes observed in the experiment.

\section*{Supplementary Information}
\subsection*{Simulation of microwave dressed-state potentials}
Here we describe our simulation of the trapping potential \cite{TreutleinThesis08}, which extends the theory of \cite{Agosta89} to multi-level atoms.
Consider a $^{87}$Rb atom in the $5S_{1/2}$ ground state at a fixed position $\mathbf{r}$. The atom interacts with the local static magnetic field $\mathbf{B}$ and the microwave magnetic field $\mathbf{B}_\mathrm{mw}(t)=\tfrac{1}{2}[\mathbf{B}_\mathrm{mw}e^{-i\omega t} + \mathbf{B}_\mathrm{mw}^*e^{i\omega t}]$. The atomic hyperfine states are described by the Hamiltonian
\[
H = (\hbar \omega_\mathrm{hfs}/2) \mathbf{I}\cdot\mathbf{J}
+ \mu_B \left(g_J \mathbf{J} + g_I \mathbf{I}\right)\cdot \left(\mathbf{B}
+ \mathbf{B}_\mathrm{mw}(t)\right).
\]
The first term describes the hyperfine coupling between electron spin $\mathbf{J}$ and nuclear spin $\mathbf{I}$ ($J=1/2$, $I=3/2$, and $\omega_\mathrm{hfs}/2\pi = 6.834\,682\,611$~GHz). The second term is the coupling to the static and microwave magnetic fields, with $g_J=2.002\,331$ ($g_I=-0.000\,995$) the electron (nuclear) Land{\'e} $g$-factor. If high accuracy of the simulation is required, e.g.\ for comparison with spectroscopic measurements, we numerically determine the eigenstates and -energies of the full Hamiltonian $H$. However, in most cases, a number of approximations can be made: (1) We neglect the coupling of $\mathbf{I}$ to the external fields because $g_I \ll g_J$. (2) We treat the coupling of $\mathbf{J}$ to $\mathbf{B}$ perturbatively because $\mu_B B \ll \hbar \omega_\mathrm{hfs}$. (3) We transfer to a frame rotating at frequency $\omega$ and make the rotating-wave approximation, valid because $\mu_B B_\mathrm{mw}, \hbar|\Delta_0| \ll \hbar\omega$, where $\Delta_0 = \omega - \omega_\mathrm{hfs}$. With these approximations, we express $H$ in the basis $|F,m_F\rangle$ of the eigenstates of the hyperfine spin $\mathbf{F}=\mathbf{J}+\mathbf{I}$, with the quantization axis chosen along the local direction of $\mathbf{B}$:
\begin{equation}\label{eq:Happrox}
\begin{split}
H &\approx \sum_{m_2} \left( -\tfrac{1}{2}\hbar\Delta_0 + \hbar\omega_L m_2 \right) |2,m_2\rangle\langle2,m_2| \\
&\quad + \sum_{m_1} \left( \tfrac{1}{2}\hbar\Delta_0 - \hbar\omega_L m_1 \right) |1,m_1\rangle\langle 1,m_1| \\
&\quad + \sum_{m_1,m_2} \left[ \tfrac{1}{2}\hbar \Omega_{1,m_1}^{2,m_2} |2,m_2\rangle\langle 1,m_1| + \textrm{c.c.} \right].
\end{split}
\end{equation}
We have approximated $g_J\approx 2$ and introduced the Larmor frequency $\omega_L = \mu_B B/2\hbar$. The microwave couples the transition $|1,m_1\rangle \leftrightarrow |2,m_2\rangle$ with Rabi frequency
\[
\Omega_{1,m_1}^{2,m_2} = (2 \mu_B/\hbar) \langle 2,m_2| \mathbf{B}_\mathrm{mw}\cdot\mathbf{J} |1,m_1 \rangle
\]
and detuning
\[
\Delta_{1,m_1}^{2,m_2} = \Delta_0 - (m_2 + m_1)\omega_L.
\]

We numerically diagonalize $H$ in equation~(\ref{eq:Happrox}) to determine the local dressed states $\ket{\bar n}$ and dressed state energies $E(n)$ of the atom at position $\mathbf{r}$, where $n = 0\ldots 7$ is a suitable labelling of the states.
For position-dependent fields $\mathbf{B}(\mathbf{r})$ and $\mathbf{B}_\mathrm{mw}(\mathbf{r},t)$, an energy landscape $E(n,\mathbf{r}$) is calculated in this way.
In our experiment, the microwave dressing field is always turned on adiabatically with respect to the internal-state dynamics. The bare qubit states $\ket{0}$ and $\ket{1}$ smoothly transform into the dressed states $\ket{\bar 0}$ and $\ket{\bar 1}$ with potentials $V_{\ket{\bar{0}}}(\mathbf{r})=E(0,\mathbf{r})-\tfrac{1}{2}\hbar\Delta_0$ and $V_{\ket{\bar{1}}}(\mathbf{r})=E(1,\mathbf{r})+\tfrac{1}{2}\hbar\Delta_0$, respectively.
If the motion of the atomic wave function in these potentials is sufficiently slow, as in our experiments, the internal state follows the spatially varying fields adiabatically, and motion-induced transitions between different states $|\bar n\rangle$ are suppressed.

If the microwave is far detuned from all transitions, $|\Delta_{1,m_1}^{2,m_2}|^2 \gg |\Omega_{1,m_1}^{2,m_2}|^2$, the microwave coupling can be treated perturbatively. Each dressed state $|\bar n\rangle$ can then be identified with one of the bare states $|F,m_F\rangle$, and each transition connecting to $|F,m_F\rangle$ contributes a potential $\pm \hbar |\Omega_{1,m_1}^{2,m_2}|^2/4 \Delta_{1,m_1}^{2,m_2}$, where the $+$ ($-$) sign is for $F=1$ ($F=2$) \cite{Treutlein06b}. This is the regime relevant to our experiment. In the main text of our paper, $\Omega_R\equiv \Omega_{1,-1}^{2,-1}$ and $\Delta\equiv \Delta_{1,-1}^{2,-1}$.
Figure \ref{fig:mwISOlines} shows the spatial dependence of $|\Omega_R(\mathbf{r})|$, calculated for an ideal CPW mode.

In addition to $V_{\ket{\bar n}}(\mathbf{r})$, our simulation includes several contributions to the potential which are identical for all states of the ground state hyperfine manifold: the quasi-electrostatic potential $V_e(\mathbf{r}) = -(\alpha_0/4) |\mathbf{E}(\mathbf{r})|^2$ due to the electric field of the microwave \cite{Engler00}, where $\mathbf{E}(\mathbf{r})$ is the electric field amplitude and $\alpha_0$ is the ground state polarizability, the potential due to gravity, and the Casimir-Polder surface potential relevant at atom-surface distances below a few micrometers \cite{Lin04}.

\begin{figure}[htb]
    \centering
        \includegraphics[width=0.48\textwidth]{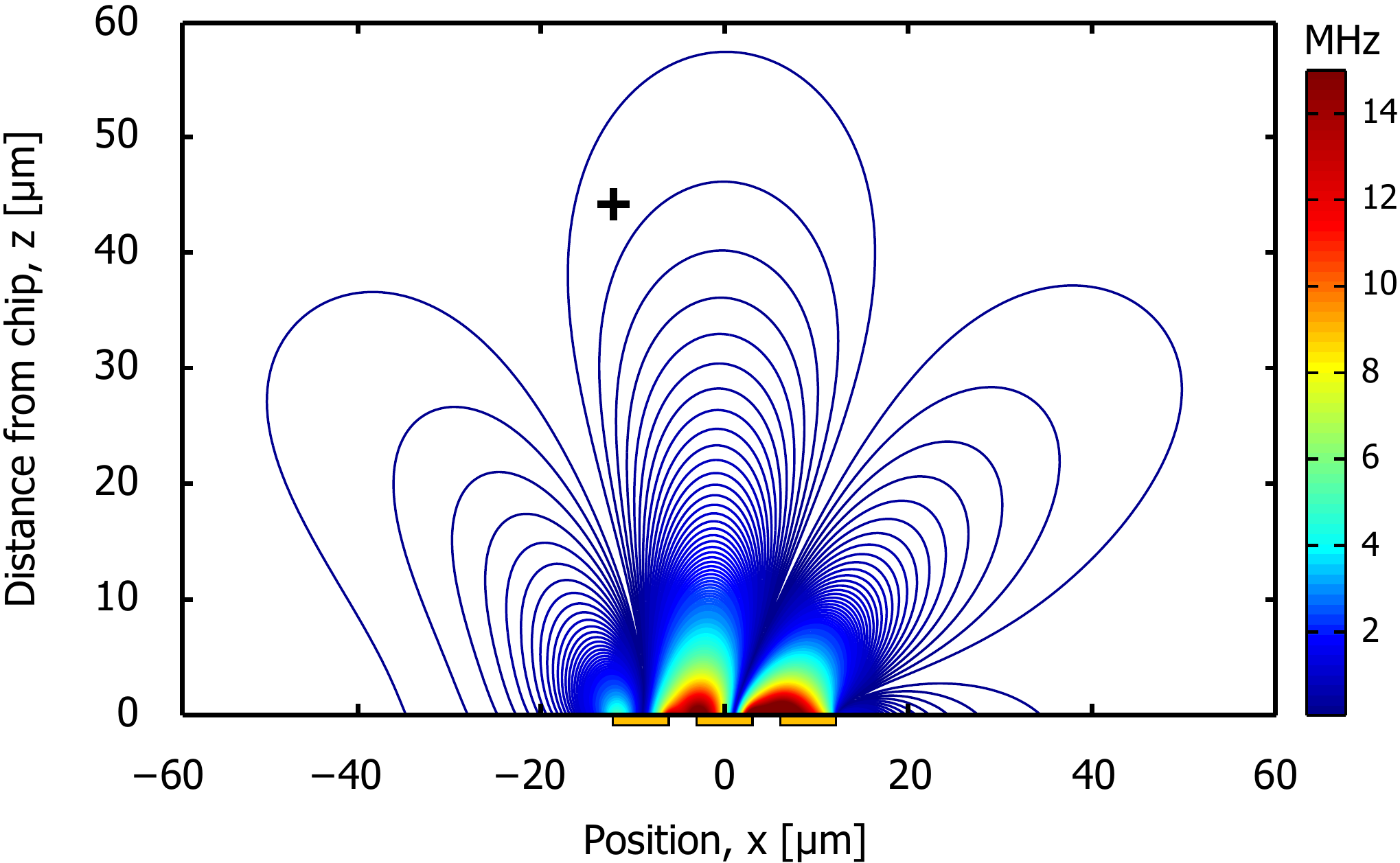}
    \caption{\textbf{Plot of the equipotential lines of $\mathbf{|\Omega_R|}$  (corresponding to Figure 2 of the main text).} Shown is $|\Omega_R(x,y_m,z)|/2\pi$, the outermost line as well as the line spacing are 70~kHz. $\Omega_R(\mathbf{r})$ is calculated from the microwave field $\mathbf{B}_\mathrm{mw}(\mathbf{r})$ of an ideal CPW mode with $\mathrm{I_{mw}=76~mA}$ and the static magnetic trapping field $\mathbf{B}(\mathbf{r})$. The asymmetry in $|\Omega_R(\mathbf{r})|$ with respect to $x=0$ is due to the spatial dependence of $\mathbf{B}(\mathbf{r})$. The position of the CPW wires is indicated below the abscissa, the trap position is marked by a black cross.
\label{fig:mwISOlines}}
\end{figure}

\subsection*{Characterization of on-chip microwave propagation}

We generate the microwave for the on-chip waveguide using an Agilent 8257D microwave generator whose output is amplified and amplitude stabilized using a feedback loop with a directional coupler and an Agilent 8471E detector directly in front of the chip. We measure a relative long-term drift in $P_\mathrm{mw}$ of $<5\cdot10^{-4}$ peak-to-peak. The relative drift of the power launched into the chip is the same as that of the transmitted power. This indicates that there is no significant long-term drift of the CPW's transmission properties. The chip temperature is stabilized by water cooling of the carrier chip.

The characteristic impedance of the CPW, obtained from quasistatic simulations including conductor and dielectric losses \cite{TreutleinThesis08}, smoothly changes from $50~\Omega$ at the ports to $70~\Omega$ at the chip center.
The measured power transmission through the chip is -6~dB. Because of symmetry of the CPW, we expect that approximately -3~dB of the injected power reach the chip center. Simulations suggest that the observed loss is dominated by conductor loss in the chip center where the lateral extent of the CPW is only a few micrometers. The absence of resonances in the transmission spectrum in the investigated frequency range ($\leq 8.5$~GHz) suggests that the bond wires and other discontinuities do not lead to strong standing waves on the chip. In any case, the atoms would experience only very small microwave potential gradients due to standing waves, because the size of the atomic wave function is much smaller than the microwave wavelength.

\section*{Acknowledgements}
We are grateful to J.~P. Kotthaus and his group for cleanroom access and advice on chip fabrication; F. Peretti, G. Csaba, F.~J. Schmückle, and F. Reinhardt for microwave simulations and helpful discussions on microwave design; and A. Sinatra and Yun Li for helpful discussions on spin squeezing. We thank S. Camerer, D. Hunger, and A. Sinatra for careful reading of the manuscript. We acknowledge support of the Nanosystems Initiative Munich.

\section*{Author information}

Correspondence and requests for materials should be addressed to P.T. (treutlein@lmu.de).

\section*{Author contributions}
P.T. and J.R. conceived the experiment.
P.T. worked out the theory.
P.B., M.R., J.H., and P.T. designed and built the experiment.
P.B., M.R., and P.T. collected and analyzed the data and wrote the manuscript.
P.T. and T.W.H. supervised the experiment.
All authors discussed the results and commented on the manuscript.

\section*{Competing financial interests}
The authors declare that they have no competing financial interests.

\end{document}